This is the accepted manuscript version of the article. The final published version is available in the *Journal of Computational and Nonlinear Dynamics* via https://doi.org/10.1115/1.4065440.

# Data-driven Initial Gap Identification of Piecewise-linear Systems using Sparse Regression and Universal Approximation Theorem

**Ryosuke Kanki and Akira Saito***
Meiji university
Kawasaki, Kanagawa 214-8571, Japan
Email: asaito@meiji.ac.jp

## Abstract

*This paper proposes a method for identifying an initial gap in piecewise-linear systems from data. Piecewise-linear systems appear in many engineered systems such as degraded mechanical systems and infrastructures, and are known to show strong nonlinearities. To analyze the behavior of such piecewise-linear systems, it is necessary to identify the initial gap, at which the system behavior switches. The proposed method identifies the initial gap by discovering the governing equations using sparse regression and calculating the gap based on the universal approximation theorem. A key step to achieve this is to approximate a piecewise-linear function by a finite sum of piecewise-linear functions in sparse regression. The equivalent gap is then calculated from the coefficients of the multiple piecewise-linear functions and their respective switching points in the obtained equation. The proposed method is first applied to a numerical model to confirm its applicability to piecewise-linear systems. Experimental validation of the proposed method has then been conducted with a simple mass-spring-hopping system, where the method successfully identifies the initial gap in the system with high accuracy.*

## 1 Introduction

It is important to understand the characteristics of complex mechanical systems to predict and control their dynamical behavior. Among such systems, piecewise-linear (PWL) systems are known to have strong nonlinearities and it is difficult to predict and analyze their dynamical behavior [1]. For instance, with the presence of joints or cracks in a mechanical structure, the system has different vibration characteristics from the ones expected in the design phase without considering its nonlinearity, causing unexpected behavior of the system [2]. In particular, intermittent contact occurs at joints or cracks due to internal vibration of the mechanical system or external forced vibration. During a vibration cycle, a PWL system has multiple linear states: the one with contact and the one without contact. The system that combines such multiple linear states is called a PWL system, and has nonlinearity due to state

* Addresses all correspondence to this author.

switching [3,4]. To understand the behavior of a PWL system, it is necessary to measure the *initial gap* that exists in the system, which is the key parameter of a PWL system because that determines where the system switches its state. In this paper, the gap is defined as the distance from the equilibrium state position to the point where the behavior switches in a PWL system. Let us say the gap of an object whose dynamics is represented by a PWL system is unknown. To analyze the dynamical behavior of the object, it needs to be disassembled and the dynamics should directly be measured. Since this involves costly and time-consuming processes, there is a strong need to develop a method to obtain information on gaps from data measurements.

Recent developments in sensing technology and data-driven modeling methods have attempted to automatically generate governing equations that describe systems from measured data [5-10]. Brunton et al. proposed a method called sparse identification of nonlinear dynamics (SINDy) for obtaining accurate governing equations from data [5]. The method uses *sparse* regression, which is an application of successive multiple regression analyses with leaving only the coefficients of the terms of high importance by setting a threshold on the regression results. Then, regression is performed again with active coefficients, which enables accurate equation derivation with sparse set of terms. This paper proposes a method for identifying an initial gap in PWL systems based on the sparse regression. To approximate the PWL elements of the measurement target, multiple PWL terms are introduced into the list of functions called a *library* and are used to perform sparse regression. Then, the equivalent stiffness and the equivalent gap are calculated from the PWL terms and their coefficients used in the approximation. With the proposed method, since the governing equation is discovered from data, additional parameter identifications such as stiffness and damping are not necessary, which is needed for classical model-based gap identification methods.

The remainder of this paper is organized as follows. In Chapter 2, mathematical foundation of the proposed method for the initial gap identification is described. In Chapter 3, initial gap identification from numerically generated data of two PWL models are provided. In Chapter 4, initial gap identification from realistic model using experimental setup is shown. In Chapter 5, concluding remarks are provided.

## 2 Method

### 2.1 Problem definition

The goal of the method is to identify an initial gap that exists in a PWL system. The PWL systems are defined as dynamical systems whose linear stiffness changes depending on the state of the variables that constitute the dynamical system, which is studied by many researchers [11-13]. PWL systems are known to exhibit complex dynamical behavior due to the switching between multiple linear states [1].

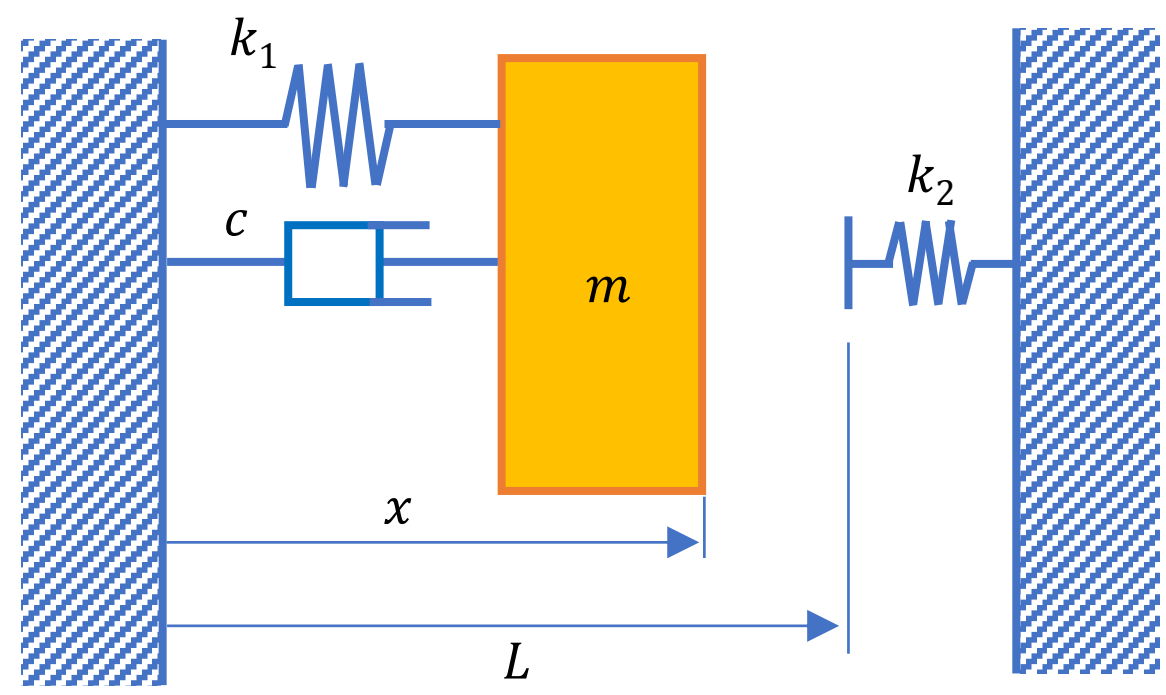

Figure 1: Mathematical model of a piecewise-linear system with an initial gap

Such systems can be found in many engineering systems, such as automobiles consisting of many components, bridges, and tunnels with cracks due to aging [14]. The PWL systems of interest in this study arise in such systems, and are assumed to be modeled as shown in Fig.1, whose equation of motion can be expressed as follows:

$$m\ddot{x} + c\dot{x} + k_1 x = -k_2 \max\{0, x - L\}, \tag{1}$$

where $m$ and $c$ are mass and damping coefficients, $k_1$ and $k_2$ are spring constants, $x$ is the displacement $L$ is the initial gap and $\max\{0, x\} = x$ for $x > 0$, or 0 for $x \leq 0$. If the mass vibrates under external disturbances, contact may or may not occur depending on its displacement $x$. The stiffness switches as the clearance repeatedly opens and closes due to the oscillation. This means that the characteristics of the structure are nonlinear, which makes the analysis of its dynamical behavior difficult. To date, there have been many attempts to develop an accurate gap identification method, and various studies have been conducted [14,15]. In this study, we propose a novel gap identification method based on the derivation of the equation of motion of the system by sparse regression. In particular, the proposed method uses a property that a PWL term, or more specifically, max function, can be approximated by a finite sum of multiple max functions, i.e.,

$$k\max\{0, x - L\} = k_1 \max\{0, x - L_1\} + k_2 \max\{0, x - L_2\} \cdots, \tag{2}$$

where $L_i$ are pre-defined gaps and $k_i$ are the corresponding spring constants. This holds because of the universal approximation theorem, i.e., it guarantees that we can find that the set of $L_i$ and $k_i$ such that Eq. (2) holds. Next, the universal approximation theorem is reviewed.

## 2.2 Universal approximation theorem applied to PWL functions

The concept of neural network has widely been used in various research fields. It is also utilized in data-driven modeling methods [16,17]. Underlying the popularity of neural networks is their universal function approximation capability. This is called the universal approximation theorem and is a theorem that ensures that any bounded continuous function can be approximated by combining nonlinear transformations of input variables [18]. In shallow neural networks, their versatility has been proven in various ways [18-21]. The proposed initial gap identification method relies on the universal approximation theorem that ensures that neural networks are universal function approximators. The proposed method implicitly uses the idea of a shallow neural network consisting of three layers: an input layer, a hidden layer, and an output layer. Hereafter, shallow neural networks are simply denoted as neural networks. In general, the approximated function $f(x) \in \mathbb{C}$ with input $x \in \mathbb{R}^m$ can be expressed using a neural network concept as follows:

$$f(x) = \sum_{j=1}^{J} c_j \eta(a_j \cdot x - b_j), \qquad (a_j, b_j, c_j) \in \mathbb{R}^m \times \mathbb{R} \times \mathbb{C}, \tag{3}$$

where $\eta$ is called the activation function, which is typically nonlinear, $a_j, b_j$ and $c_j$ are a gradient, an intercept, and an output parameter, respectively, and $J$ is the number of neurons. Equation (3) has been proved by Sonoda and Murata to hold if the function to be approximated $f(x) \in L_1, L_2$ and the activation function $\eta \in \mathcal{S}'$ [18], where $L_1, L_2$ are integrable and square-integrable function spaces respectively, i.e.,

$$\begin{aligned} &\int_{\mathbb{R}} |f(x)| \mathrm{d}x < \infty, \\ &\int_{\mathbb{R}} |f(x)|^2 \mathrm{d}x < \infty, \end{aligned} \tag{4}$$

$\mathcal{S}'$ is the space containing the tempered distributions. Since the system targeted in this study deals only with bounded intervals, the approximated function $f(x)$ satisfies these conditions. Now we attempt to apply this to the PWL term in Eq. (1). Namely, we assume that the activation function $\eta$ is a max function in this case. It is known that the max function can be used as the activation function [21]. From the above, if the possible range of $x$ is a bounded interval and the function to be approximated is *also* a max function, the following is expected to hold, given appropriate $k_j, L_j (\neq L)$,

$$k \max\{0, x - L\} \approx \sum_{j=1}^{J} k_j \max\{0, x - L_j\}, \tag{5}$$

where $J$ is the number of terms to be used in the approximation. This means that a PWL function can be approximated by a finite sum of multiple PWL functions.

By utilizing Eq. (5), the equivalent stiffness and equivalent gap of the PWL system can be obtained. When $x \leq L$, the PWL term vanishes in Eq.(1). On the other hand, when $x > L$ on the left-hand side of Eq. (5), the equation can be expressed as:

$$k\max\{0, x - L\} = kx - kL. \tag{6}$$

Next, considering the right-hand side of Eq.(5). If, $L_n < x < L_{n+1}$, for $n + 1 < J$, the right-hand side of Eq. (5) can be expressed as:

$$\sum_{j=1}^{J} k_j \max\{0, x - L_j\} = \sum_{j=1}^{n} k_j \max\{0, x - L_j\}$$

$$= x \sum_{j=1}^{n} k_j - \sum_{j=1}^{n} k_j L_j \,. \tag{7}$$

From Eq. (5), since Eq. (6) approximately equals to Eq. (7), the terms on the right hand side of Eq.(7) are good approximations of the corresponding terms of Eq.(6). Namely, by comparing the coefficients for $x$ on both sides, the following relationship holds:

$$k = \sum_{j=1}^{n} k_j \,. \tag{8}$$

Similarly, from the constant terms in Eqs. (6) and (7), the relationship of $kL = \sum_{j=1}^{n} k_j L_j$ holds. Therefore, the gap can be determined as:

$$L = \frac{1}{k} \sum_{j=1}^{n} k_j \, L_j. \tag{9}$$

From Eqs. (8) and (9), the equivalent spring constant $k_{eq}$ and the equivalent gap $L_{eq}$ are defined as follows:

$$k_{eq} := \sum_{j=1}^{n} k_j \,, \quad L_{eq} := \frac{1}{k_{eq}} \sum_{j=1}^{n} k_j L_j \,. \tag{10}$$

The relationship in Eq. (5) holds even if the max function becomes the min function. If the assumed spring force is described by a min function, the right-hand side of Eq. (5) is changed to the sum of min functions. This allows the relationship of Eq. (10) to be utilized in a min function also, as it will be shown in 3.2.

## 2.3 Sparse regression with PWL functions

In section 2.2, universal approximation theorem was utilized to approximate the PWL function, and the equivalent spring constant and gap were introduced. In this section, the sparse regression method proposed in Ref. [5] is extended such that it can be applied to PWL systems. The state equation under consideration is designated as $\dot{\mathbf{X}} = \mathbf{f}(\mathbf{X}(t))$ where $\mathbf{X} \in \mathbb{R}^{p \times q}$ is a matrix containing the time histories of the state variables, $p$ is the number of time instants, $q$ is the number of states, and $\mathbf{f}$ denotes the governing equation that describes the behavior of the dynamical system. It is assumed that the data is collected at time instants $t = [t_1, \cdots, t_p]$. First, $\mathbf{X}$ that consists of time series data such as displacement and velocity, and $\dot{\mathbf{X}}$ that consists of the time derivatives of the data such as velocity and acceleration are created. Next, we construct $\mathbf{\Theta}(\mathbf{X})$, which contains the time history of the group of the candidate functions. $\mathbf{\Theta}(\mathbf{X})$ is called a library and consists of a constant and polynomial terms. The SINDy identifies the equation by discovering the relationship between $\dot{\mathbf{X}}$ and $\mathbf{\Theta}(\mathbf{X})$ using the thresholded least squares method, leaving only the important variables.

In applying SINDy to the PWL systems of interest, we propose that multiple PWL functions with different gap length be included in the library. Namely, assuming that $\mathbf{X} = [\mathbf{v}, \quad \mathbf{x}] \in \mathbb{R}^{p\times 2}$,

$$\mathbf{\Theta}(\mathbf{X}) = \begin{bmatrix} | & | & | & | & & | & & | \\ \mathbf{1} & \mathbf{X} & \mathbf{P}_2(\mathbf{X}) & \mathbf{P}_3(\mathbf{X}) & \cdots & \max\{0, \mathbf{x} - L_1\} & \cdots & \max\{0, \mathbf{x} - L_n\} \\ | & | & | & | & & | & & | \end{bmatrix}. \tag{11}$$

Here, higher polynomials are denoted as $\mathbf{P}_2(\mathbf{X}), \cdots, \mathbf{P}_O(\mathbf{X})$. For instance, $\mathbf{P}_2(\mathbf{X}) = [\mathbf{v}^2 \quad \mathbf{vx} \quad \mathbf{x}^2]$ In this study, the order of the library is designated as $O$. For instance, if $O = 3$, the library includes the time histories of polynomials up to $\mathbf{P}_3$. With these variables, the following regression problem is defined:

$$\dot{\mathbf{X}} = \mathbf{\Theta}(\mathbf{X})\mathbf{\Xi}. \tag{12}$$

where $\mathbf{\Xi} = [\xi_1 \quad \cdots \quad \xi_N]$ are the coefficients to be determined using the thresholded least squares method. In sparse regression, a threshold is set after a polynomial approximation by general least squares method. Namely, the vectors of coefficients that are smaller than the threshold are forced to be zero and the remaining variables are used to solve another least squares problem with the fewer variables. More specifically, denoting the threshold as $\lambda$, the columns in $\mathbf{\Theta}(\mathbf{X})$ corresponding to $i$ that satisfies $\lambda > |\xi_i|$ are eliminated from the library. This leaves only the necessary terms to describe the behavior of the system in the coefficient vector, or the system coefficients become sparse. This process is continued until the convergence of $\mathbf{\Xi}$ is achieved. The equations that approximates the original model can be derived as follows:

$$\dot{\mathbf{X}}(t) = \mathbf{f}\big(\mathbf{X}(t)\big) \approx \mathbf{\Theta}(\mathbf{X})\mathbf{\Xi}_{\mathrm{sparse}}. \tag{13}$$

where $\mathbf{\Xi}_{\mathrm{sparse}}$ is the sparse set of coefficients corresponding to the functions in the library.

In summary, the relationship Eq. (5) is used to approximate the PWL element and to obtain the equivalent gap from the discovered equations. Based on the sparse regression method, the governing equation is then derived using Eq. (13), the equivalent gap is identified by applying Eq. (10). The advantage of the proposed approach over the existing gap identification method is that it does not require the governing equation of the system a priori. For instance, the method proposed in Refs. [14, 15] requires the development of a mathematical model of the system before the application of the gap identification method. Furthermore, it requires the measurement or identification of parameters such as stiffness and damping after the model is built. On the other hand, the proposed method only requires the time histories of the displacement. Moreover, the method identifies not only the gap but also other system parameters such as stiffness and damping simultaneously during the regression.

It is also noted that the method can approximate nonlinear contact forces that arise in general PWL systems even if the contact force involves arbitrary contact stiffness in each of its linear states with multiple stoppers, for instance. However, it is not possible to identify the gaps of multiple max functions, because it is not yet possible to properly distribute the contributions of the max functions in the library to the multiple max functions that indeed exist in the system. To handle such cases, the extension of the method is necessary.

The extension of the proposed approach to multiple DOF cases with multiple contacting masses is not straightforward and beyond the scope of this paper. However, by properly choosing the form of max functions in the library, it may be able to handle multiple DOF cases. If that becomes possible, the extension of the method to even flexible structures with contact boundaries should not be difficult. It is because the solution of the governing equations of flexible structures typically involves the spatial discretizations such as finite element method, which results in the discrete dynamical system that can be handled by the proposed approach.

## 2.4 Numerical examination of the equivalent spring constant and gap

It was shown in section 2.2 that Eq. (5) holds and the equivalent spring constant and gap can be identified by Eq. (10). In this section, validity of these relationships is examined based on numerical examples. To illustrate that a max function can be described by a finite sum of multiple max functions, single max function $F = k\max\{0, x - L\}$ is approximated using five different max functions in this example as follows:

$$k\max\{0, x - L\} \approx \sum_{j=1}^{5} k_j \max\{0, x - L_j\}. \tag{14}$$

In this numerical example, $k = 1$ and $L = 1.5$ in $F$. The gaps of the candidate PWL functions are $L_j = (j - 1)L_0$, $j = 1, \cdots, 5$, and $L_0 = 1.0$. The coefficients $k_j$ of the terms are found using a standard least-squares method. Two types of dataset of $x$ were prepared as follows:

Case (A): $$x = 10\sin(\omega t) \tag{15}$$

Case (B): $$x = 10\sin(\omega t) + 5\cos(3\omega t), \tag{16}$$

where $\omega = 1$ [rad/s]. $t$ was set from 0 to 10 [s], and the time step was $\Delta t = 0.01$[s]. The left-hand side of Eq. (13) contains the time-series data of $F$. On the right hand side of Eq. (13), time-series data of multiple candidate max functions are placed in $\mathbf{\Theta}$.

Table 1 shows the coefficients assigned to each max function by the least squares method. Table 2 shows the equivalent stiffness and equivalent gap calculated from the values in Table 1 using the relationship in Eq. (10). The values of $k_{eq}$ and $L_{eq}$ are expected to be close to 1 and 1.5, respectively. From Table 2, in Case (A), the spring constant and the gap were accurately identified with $0.02\%$ and $0.14\%$ errors, respectively. In Case (B), the errors of the spring constant and the gap were $0.01\%$ and $0.06\%$, respectively. From both results, we can see that the equivalent spring constant and equivalent gap can be obtained quite accurately with small errors. Therefore, Eqs. (5) and (10) are found to be numerically valid. The trajectories of $F$ for both cases are shown in Fig. 2. The solid line is the time history of the max function to be approximated, and the dotted lines are the time histories of the individual max functions, and the dashed line is the linear combination of the time histories of the individual max functions. As can be seen, the time history computed by the linear combination fits the original trajectory quite well. In both Cases (A) and (B), there are good agreements between the time history of the function to be approximated and that obtained by the regression. Therefore, it can be visually understood that Eq. (5) holds. From these results obtained, we can see that the accurate gap identification is possible by the proposed approach.

Table 1: Coefficients assigned to each max function.

| | $k_1$ | $k_2$ | $k_3$ | $k_4$ | $k_5$ |
|---|---|---|---|---|---|
| Case(A) | -0.0741 | 0.5686 | 0.6051 | -0.1211 | 0.0217 |
| Case(B) | -0.0753 | 0.5724 | 0.5878 | -0.0921 | 0.0073 |

Table 2: Equivalent spring constant, equivalent gap and their errors.

| | $k_{eq}$ | $L_{eq}$ | Error of $k_{eq}$[%] | Error of $L_{eq}$[%] |
|---|---|---|---|---|
| Case (A) | 1.0002 | 1.5022 | 0.02 | 0.14 |
| Case (B) | 1.0001 | 1.5009 | 0.01 | 0.06 |

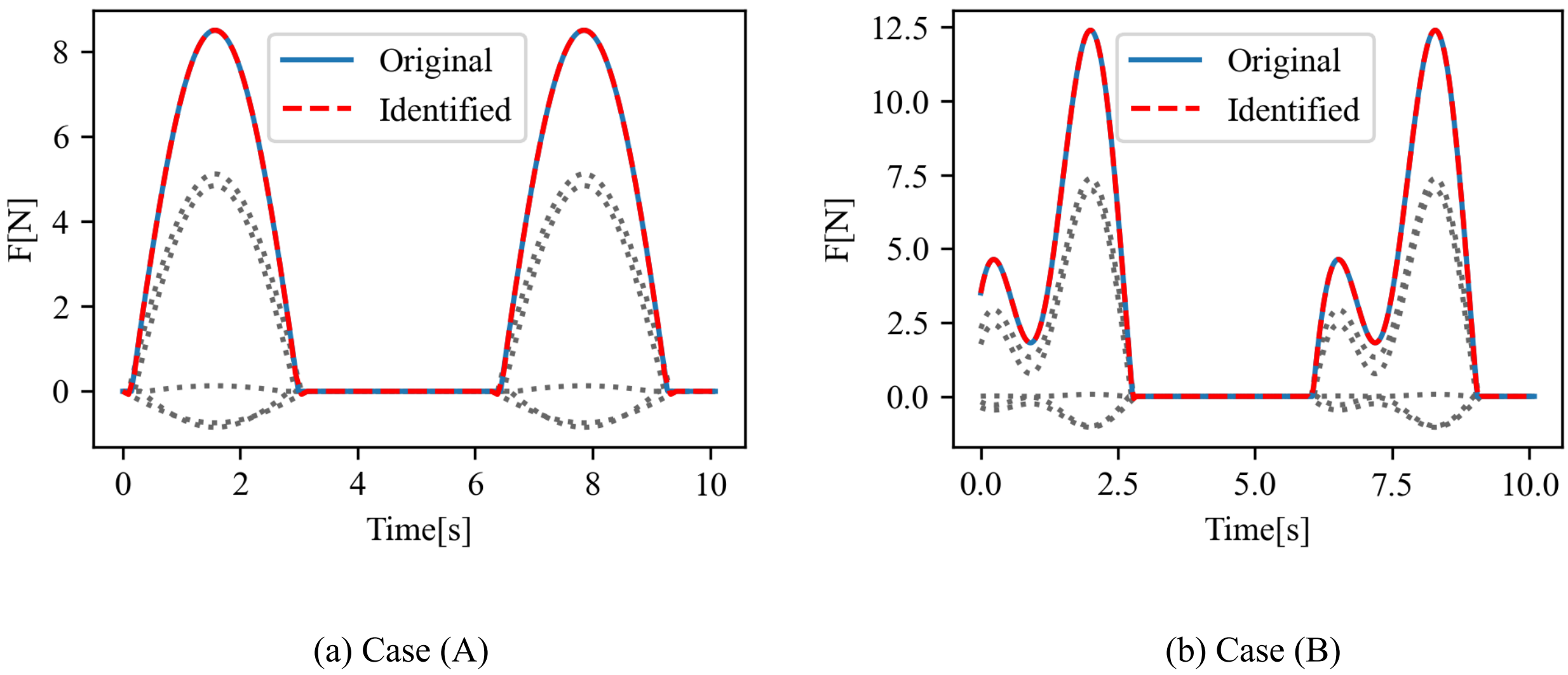

(a) Case (A)

(b) Case (B)

Figure 2: Trajectories of original data and equations obtained by regression.

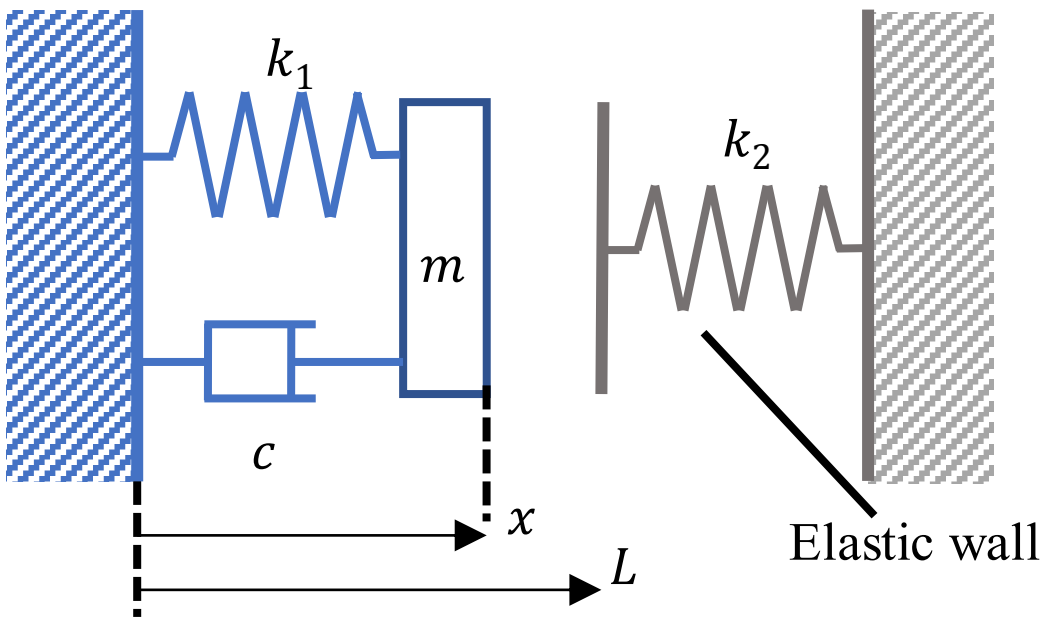

(a) The mass-spring-damper system with an elastic wall

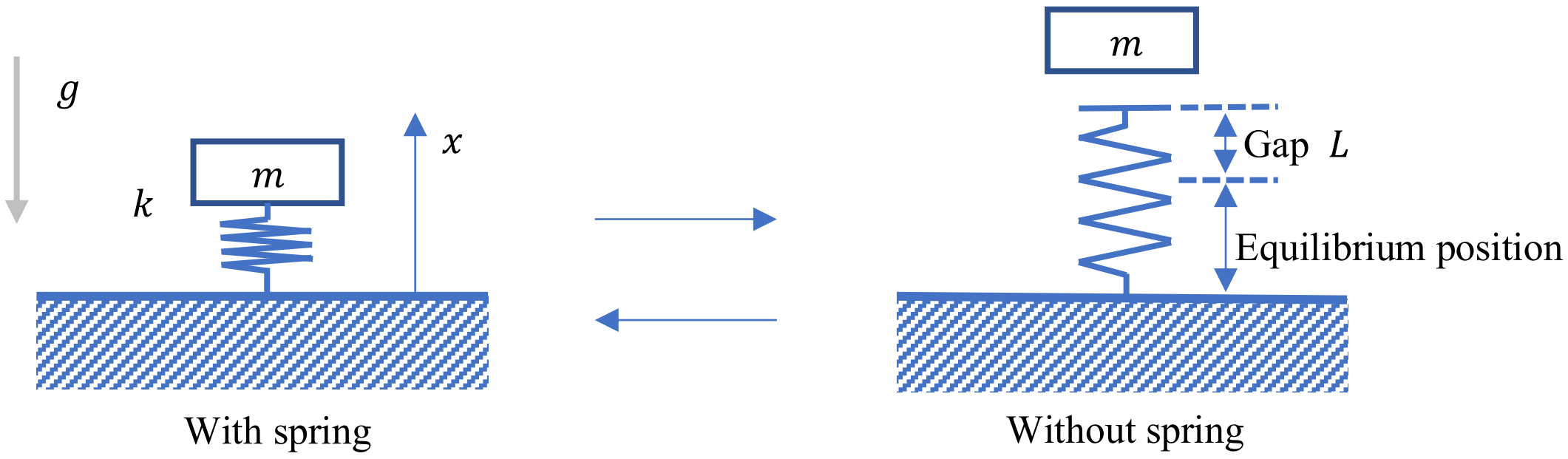

(b) The mass-spring hopping system

Figure 3: The target of two types of PWL systems.

## 3 Numerical analyses

In this section, the gap identification of a PWL system using the proposed method is applied to numerical models. In the numerical analysis, gap identification is performed for two types of systems.

### 3.1 PWL Mass-spring-damper system

The first numerical example is a PWL mass-spring-damper system in which the mass is subject to intermittent contact with an elastic wall, as shown in Fig.3(a). The mass is given an initial displacement and velocity, and undergoes free vibration. When the displacement of the mass exceeds a certain point, the mass contacts with the wall. The stiffness then switches, and the behavior changes. Therefore, the governing equations for the system in Fig. 3(a) can be expressed as follows:

$$m\ddot{x} + c\dot{x} + k_1 x = -k_2 \max\{0, x - L\}, \tag{17}$$

where $m, c,$ and $k_1$ are mass, damping, and spring constant. $k_2$ is the spring constant of the wall and $L$ is the distance from the equilibrium position of the mass to the elastic wall. The initial gap, which in this case is $L$, is identified from the model shown in Fig. 3(a) using the proposed method. Eq.

(17) is transformed to a state-space representation, as follows:

$$\begin{aligned} \dot{v} &= -\frac{c}{m}v - \frac{k_1}{m}x - \frac{k_2}{m}\max\{0, x - L\} \\ \dot{x} &= v \end{aligned} \tag{18}$$

The conditions for the analyses were set as follows. The displacement, velocity, and acceleration were assumed to be measured and available. In the numerical experiments, displacement and velocity data for the model shown in Fig. 3 were generated by a numerical time integration based on Eq. (18). The `odeint` function available in Python's Scipy library was used for the numerical integration. The acceleration data were generated by substituting these displacement and velocity data into Eq. (18). Parameters were set to $m = 5$, $c = 10$, $k_1 = k_2 = 100$, and $L = 1.5$. The initial states were set to $[v(0), \quad x(0)] = [0.01, \quad -0.1]$. $\mathbf{X}$ and $\dot{\mathbf{X}}$ were collected in the interval $t = 0$ to 10 [s] at time increment $\Delta t = 0.001$ [s]. The threshold for sparse regression was set to $\lambda = 0.05$. In the library $\mathbf{\Theta}$, polynomial terms up to $O = 3$ were considered, and the number of candidate max functions was set to $n = 5$. The gap $L_j$ for each max function was chosen to be $L_j = (j-1)L_0$, $L_0 = 1.0$, $j = 1, \cdots, n$. The solution obtained by the numerical integration was used as the data that was supposed to be measured. Furthermore, to investigate the effects of measurement noise, artificial measurement noise was injected to the data by a Gaussian noise with zero mean and standard deviation of $1.0\%$ of the maximum displacement. Figure 4 shows the time histories used for identification.

The obtained coefficients of the terms, the equivalent stiffnesses and the identified gaps are shown in Table 3 along with the correct values to be identified. The coefficients expressed in the table are shown to the third decimal place. Terms with zero derived coefficients are omitted. In the resulting equations, terms that were not present in the original equations, with or without noise, became zero. Not surprisingly, this means that the terms in the library that were not in the original equations do not contribute to the dynamics. The identified equivalent stiffness without noise has an error of $0.01\%$ and that with noise is $0.015\%$. The identified equivalent gap also has an error of $0.13\%$ for the case without noise and $0.40\%$ with the noise. Both equivalent stiffness and equivalent gap were successfully identified with small errors. Figure 5 shows a phase portrait of the solutions of the original equation and the obtained equation. As expected, the obtained solution trajectories are in good agreement with the solution trajectory of the original equation.

Table 3: The identified coefficients of the equation of motion of mass-spring-damper system.

| $\boldsymbol{\Theta}$ | Correct value | Without noise | With noise |
|---|---|---|---|
| 1 | 0.000 | 0.000 | 0.000 |
| $x$ | 20.000 | 20.000 | 20.001 |
| $v$ | 2.000 | 2.000 | 2.000 |
| $x^2$ | 0.000 | 0.000 | 0.000 |
| $\vdots$ | $\vdots$ | $\vdots$ | $\vdots$ |
| $v^3$ | 0.000 | 0.000 | 0.000 |
| $\max\{0, x - L\}$ | 20.000 | – | – |
| $\max\{0, x - L_1\}$ | – | -0.326 | -0.808 |
| $\max\{0, x - L_2\}$ | – | 8.856 | 9.067 |
| $\max\{0, x - L_3\}$ | – | 13.612 | 15.143 |
| $\max\{0, x - L_4\}$ | – | -2.529 | -4.356 |
| $\max\{0, x - L_5\}$ | – | 0.388 | 0.958 |
| $k_{eq}/m$ | 20.000 | 20.002 | 20.003 |
| $L_{eq}$ | 1.500 | 1.502 | 1.506 |

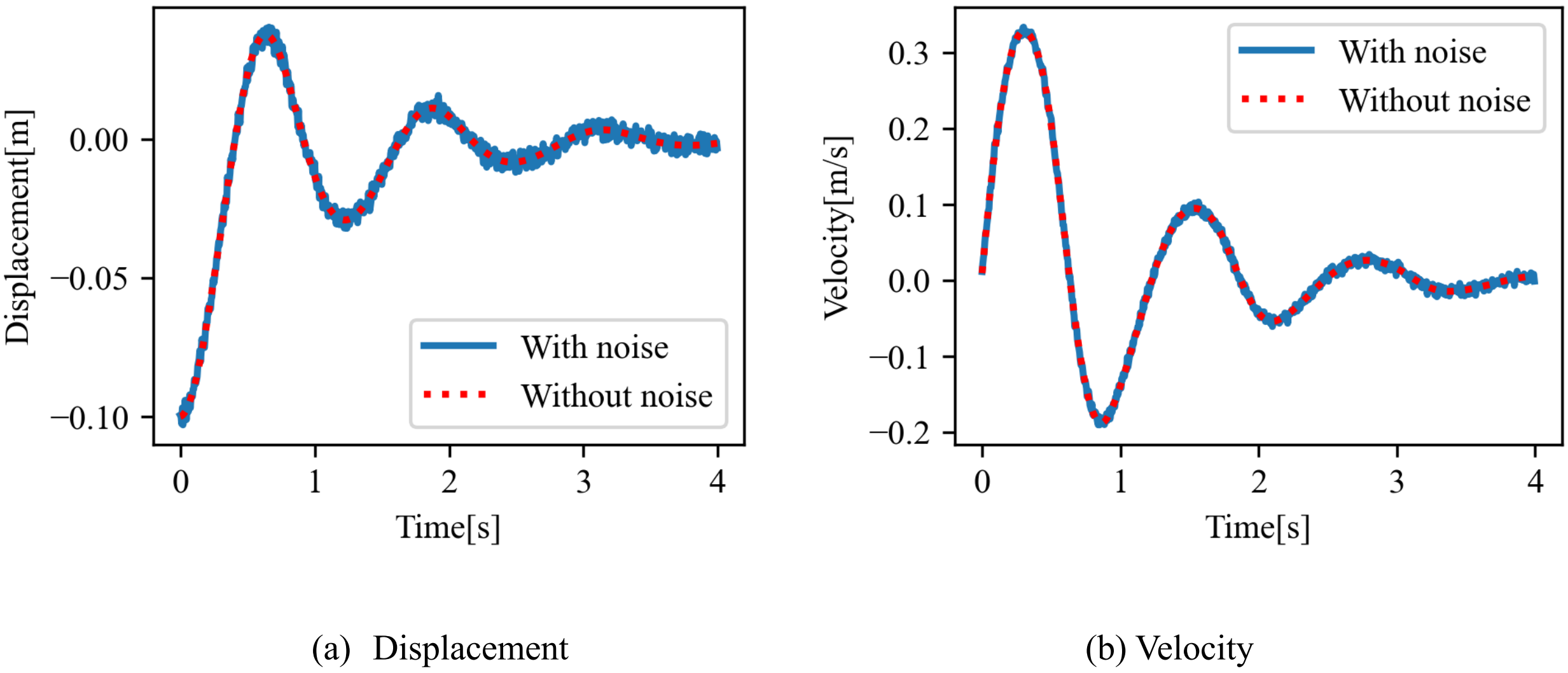

(a) Displacement (b) Velocity

Figure 4: The time histories of the displacement and velocity of the mass-spring-damper system with and without noise.

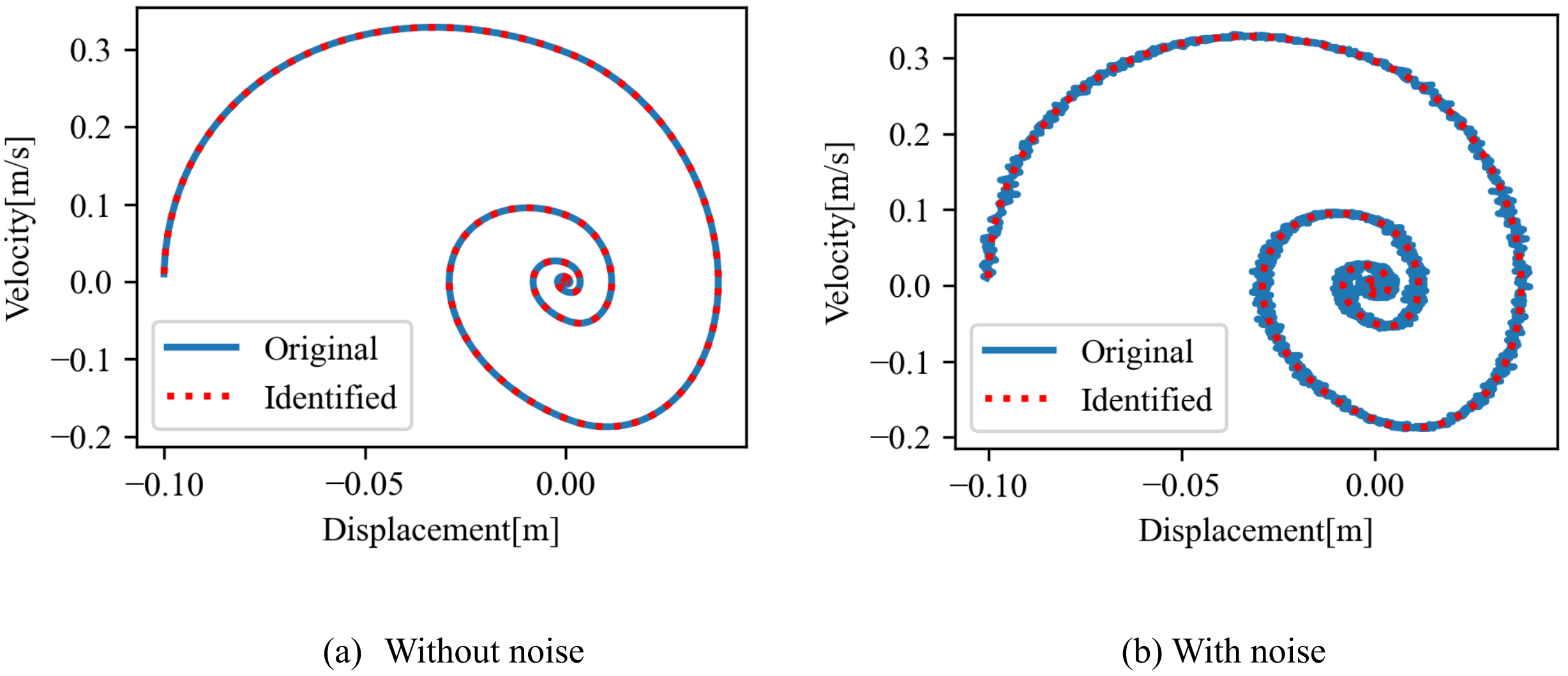

(a) Without noise (b) With noise

Figure 5: The trajectory of each PWL mass-spring-damper equation derived with and without noise.

## 3.2 Mass-spring-hopping system

The second numerical example is a mass-spring-hopping system shown in Fig. 3(b). This system has two states: one is the free vibration motion of the mass with the spring, and the other is the free fall motion of the mass. The governing equation for this system can be expressed as follows:

$$m\ddot{x} = -mg - k\min\{0, x - L\}, \tag{19}$$

where $\min\{0, x\}$ is $x$ for $x < 0$, or 0 for $x \geq 0$, $m$ and $k$ are mass and spring constants, $g$ is the gravitational acceleration. $L$ is the distance from the equilibrium position of the mass to the natural length of the spring, which is regarded as the gap in this system. The spring is fixed to the ground. Therefore, it goes back and forth between two linear states, with the initial gap $L$ as the switching point of behavior. The gap is identified from the model shown in Fig. 3(b) using the proposed method. Here, as in Section 3.1, Eq. (19) is transformed into a state space representation to apply the sparse regression.

$$\begin{aligned} \dot{v} &= -g - \frac{k}{m}\min\{0, x - L\} \\ \dot{x} &= v \end{aligned} \tag{20}$$

Denoting the gap as $L$, it can be computed as follows:

$$L = \frac{mg}{k} \tag{21}$$

The conditions for the analysis were set as follows. The displacement, velocity, and acceleration were assumed to be measured. These data were generated as in Section 3.1. Parameters were set to $g = 9.81$ [m/s$^2$], $m = 0.20$ [kg], and $k = 500$ [N/m], resulting in $L = 3.924$ [mm]. The initial states were set to $[v(0), x(0)] = [-0.01\ [\mathrm{m/s}],\ \ 0.1\ [\mathrm{m}]]$. $\mathbf{X}$ and $\dot{\mathbf{X}}$ were collected in the interval $t = 0$ to 10 [s] at time step $\Delta t = 0.001$ [s]. The threshold for sparse regression was set to $\lambda = 0.05$. In the library $\mathbf{\Theta}$, polynomial terms up to order $O = 3$ were included, and the number of candidate max functions was set to $n = 5$. The gap $L_i$ for each min function was chosen to be $L_j = (j-1)L_0$, $L_0 = 1.0$[mm], $j = 1, \cdots, n$. The used data were the solution obtained by the numerical calculation. To investigate the effects of measurement noise, artificial measurement noise was again injected to the data by a Gaussian noise with zero mean and standard deviation of 3.0% of maximum displacement. Figure 6 shows the time histories used for identification.

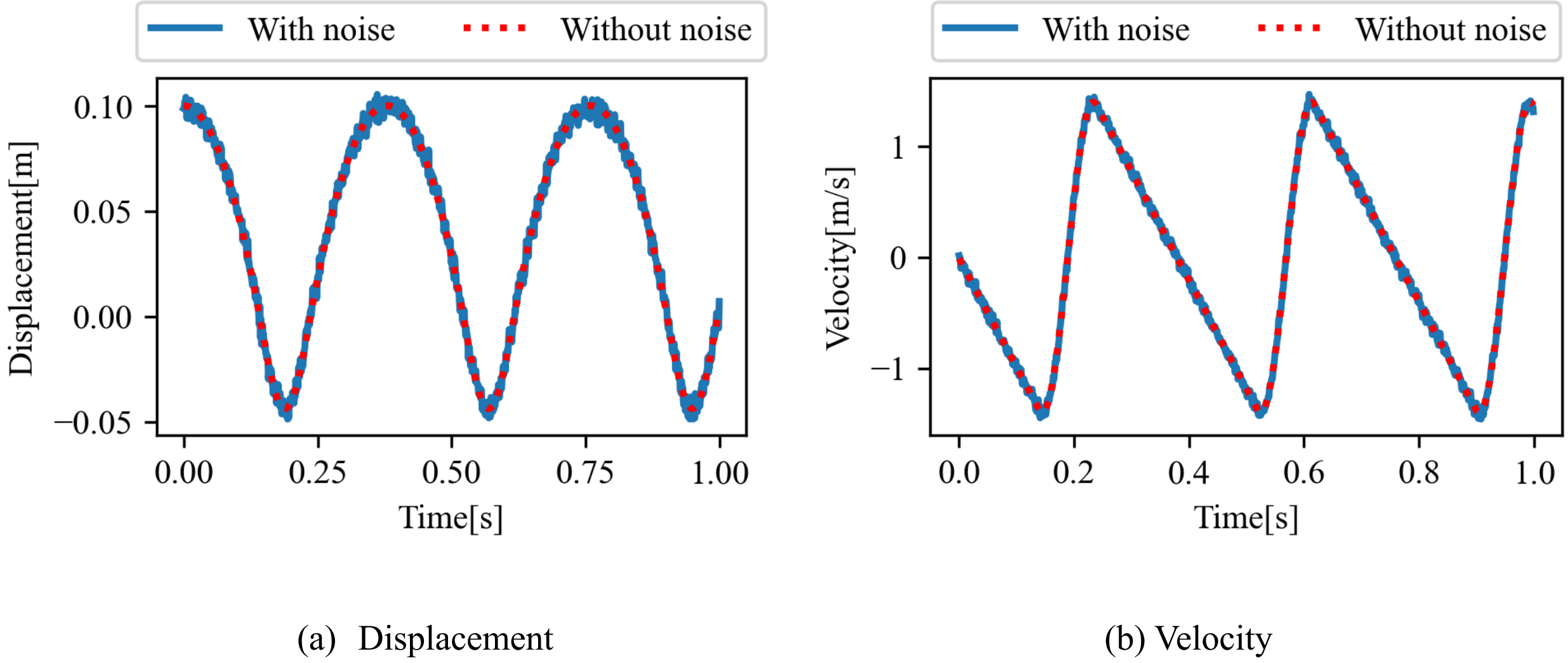

(a) Displacement (b) Velocity

Figure 6: The time histories of mass-spring-hopping system with and without noise.

The obtained coefficients of the terms, the equivalent stiffness and the gaps are shown in Table 4. As can be seen, only the terms used in the original equation appear, indicating that sparse regression was successful. The identified equivalent stiffness without noise has an error of $1.24 \times 10^{-4}$%. In the case with noise, the error is $2.22 \times 10^{-3}$%. The identified equivalent gap also has an error of $3.61 \times 10^{-3}$% for the noiseless case and $1.39 \times 10^{-2}$% for the noisy case. The identification of the equivalent stiffnesses and equivalent gaps in this example is more accurate than that shown in section 3.1. This appears to be because the PWL spring term in the equation is larger than in Section 3.1, making it relatively easier to express the characteristics. Another important feature of this example is that the PWL function used in the PWL system is the min function, i.e., the gap identification can be performed even in the case of the min function. Figure 7 shows the trajectories of the mass in the phase plane of the original equation and the obtained equation. Figure 7 shows that the obtained solution trajectories are in good agreement with each other. It shows that obtained equations captures the essence of the data, even if data contains noise.

Table 4: The identified coefficients of the equation of motion of mass-spring-hopping system.

| $\mathbf{\Theta}$ | Correct value | Without noise | With noise |
|---|---|---|---|
| 1 | 9.810 | 9.810 | 9.810 |
| $x$ | 0.000 | 0.000 | 0.000 |
| $v$ | 0.000 | 0.000 | 0.000 |
| $\vdots$ | $\vdots$ | $\vdots$ | $\vdots$ |
| $v^3$ | 0.000 | 0.000 | 0.000 |
| $\min\{0, x-L\}$ | 2500.000 | – | – |
| $\min\{0, x-L_1\}$ | – | -3.394 | -4.831 |
| $\min\{0, x-L_2\}$ | – | 24.229 | 12.638 |
| $\min\{0, x-L_3\}$ | – | -107.591 | -27.973 |
| $\min\{0, x-L_4\}$ | – | 345.717 | 225.991 |
| $\min\{0, x-L_5\}$ | – | 2241.036 | 2294.120 |
| $k_{eq}/m$ | 2500.000 | 2499.997 | 2499.945 |
| $L_{eq}$ | 3.924 | 3.924 | 3.925 |

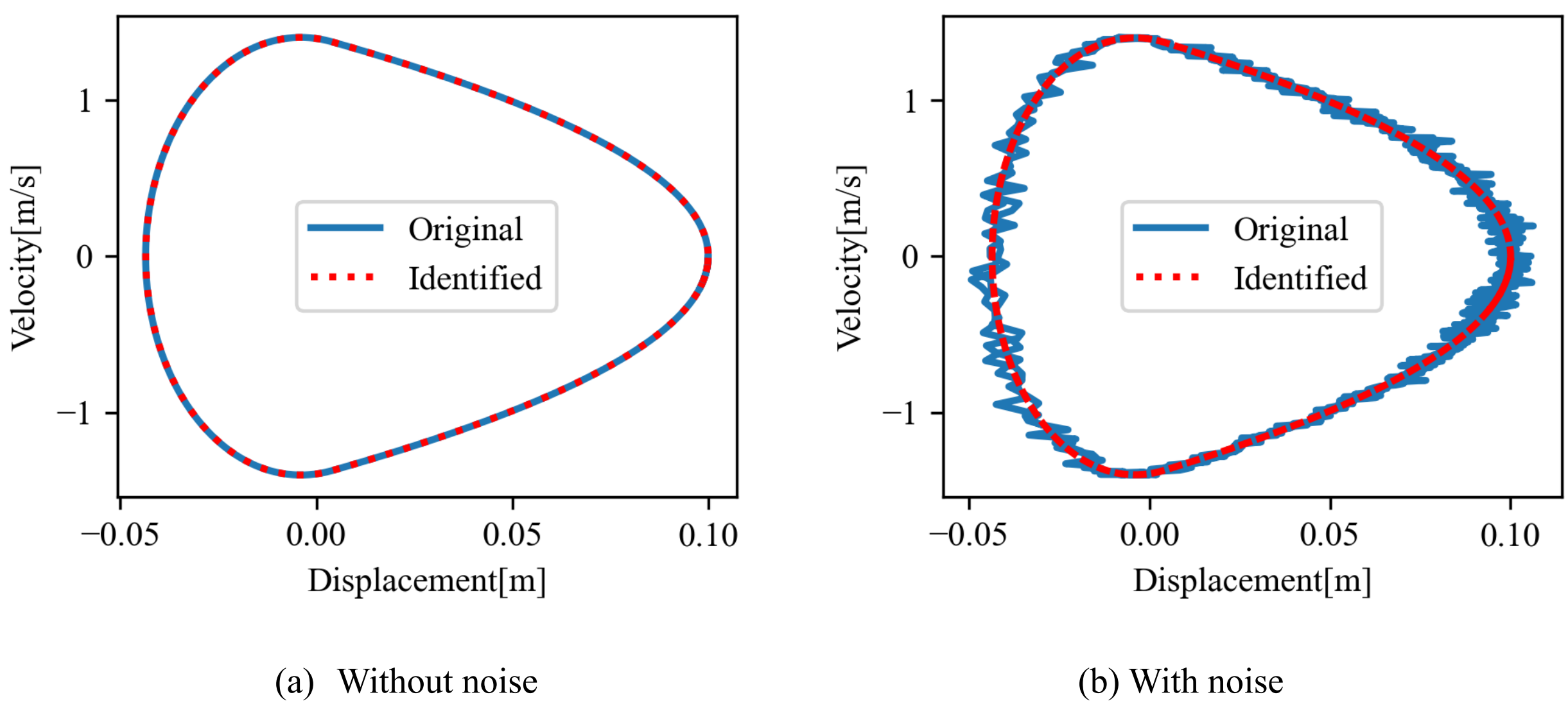

(a) Without noise (b) With noise

Figure 7: The trajectory of each equation of mass-spring-hopping system derived with and without noise.

### 3.3 Convergence study

In this section, to discuss the convergence characteristics of the proposed method with respect to the number of PWL functions, the results of a convergence study are presented with respect to the number of PWL functions in the library. The target system is the PWL mass-spring-damper system described in Section 3.1. The gaps of the candidate max functions were set from 0 to 4.0, and the number of max functions with gaps within that interval was increased. For example, if the number of terms is set to 9, the gap is divided into 9 equals parts from 0 to 4.0, or [0.0, 0.5, 1.0, 1.5, …, 4.0].

Figure 8 shows the errors in the identified equivalent spring constants and equivalent gaps, respectively. The figure indicates that the error in the identified values decreases as the number of gap divisions is increased. Note that there are several cases where the error reduces down to zero. This is because the correct max function with the gap of 1.5 is included in the library for each increment of 8 divisions. Figure 9 shows the percentage contribution of each max function for the cases of 24, 25, 50, 75, and 100 divisions. The contribution is the absolute value of the coefficient assigned to the PWL function. The solid line in each graph indicates a gap of 1.5, which is the correct solution. First, when the number of divisions is 25, there is a max function with a gap of 1.5 in the library. This sparsifies the other functions and concentrates the contribution to only one function. The other results show that the distribution of the contributions is centered around 1.5, which is the gap of the correct solution, for all divisions. The max function with a gap close to 1.5 has the highest contribution, and the further away from the correct value, the smaller the contribution becomes. The number of used max functions does not change for approximately any number of partitions, except when the library has a max function with the correct solution. In the case of 50 divisions, eight functions were assigned coefficients. In the case of 24, 75, 100, seven max functions were used.

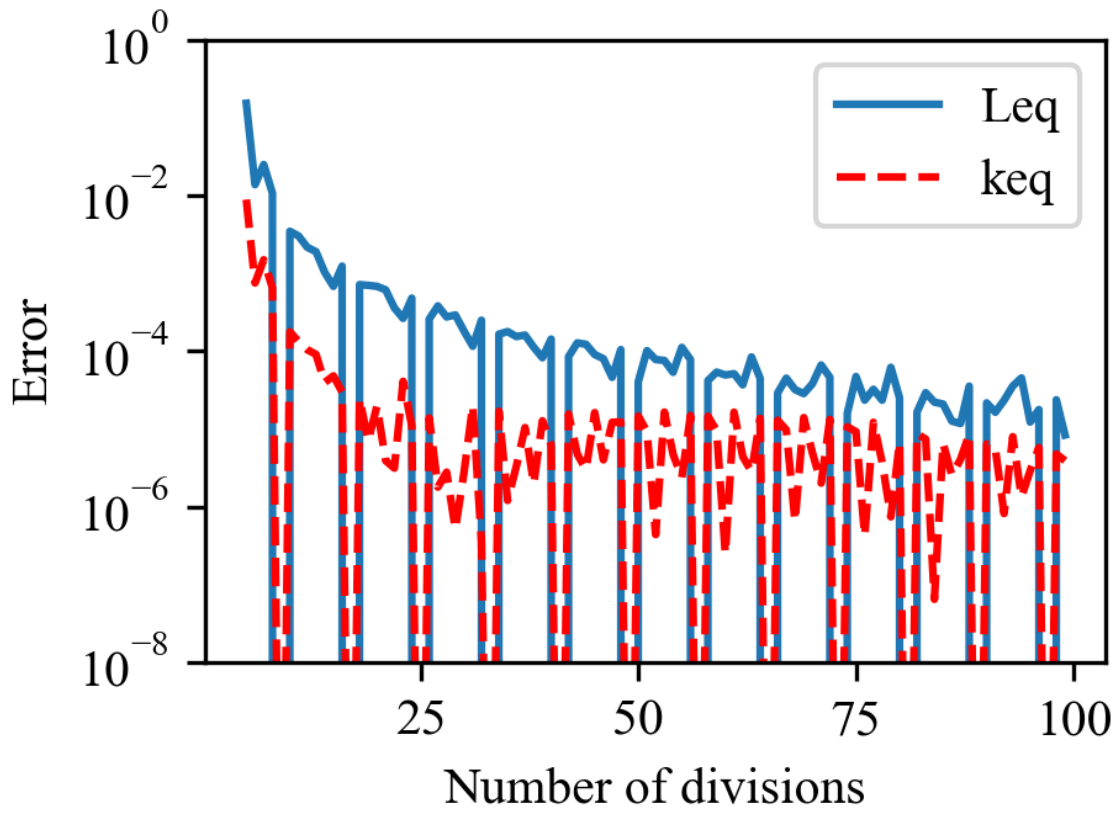

Figure 8: The error of the equivalent spring constants and the gaps for each number of divisions.

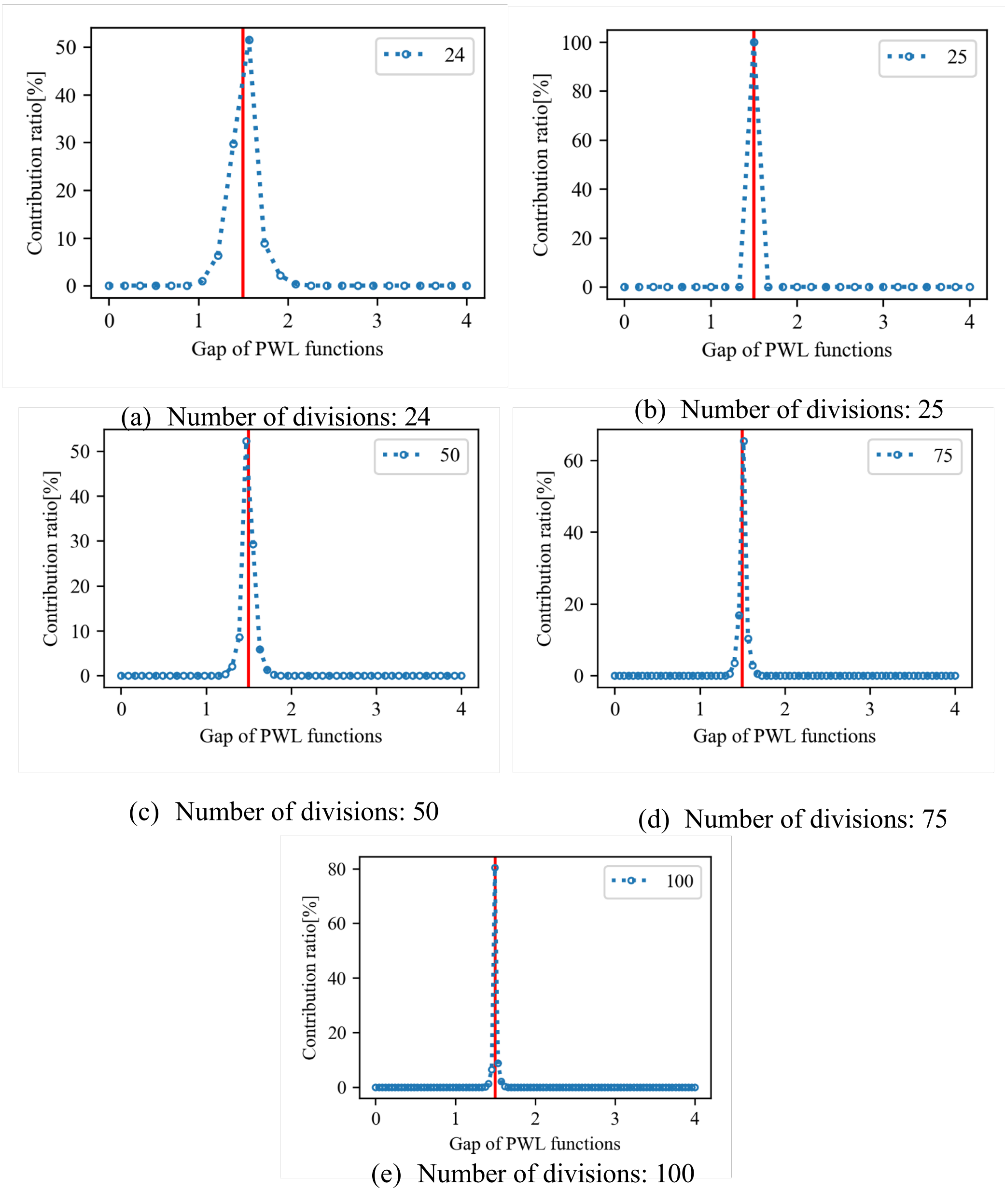

(a) Number of divisions: 24

(b) Number of divisions: 25

(c) Number of divisions: 50

(d) Number of divisions: 75

(e) Number of divisions: 100

Figure 9: Contribution of PWL functions

## 4 Experimental analysis

### 4.1 Experimental setup

This section describes the equipment used in the experimental analysis. As shown in Fig. 10, the equipment consists of a mass and a spring to represent the mass-spring-hopping system described in Section 3.2. A shaft was used to secure the mass to move vertically to the ground. Since the movement is along the shaft, the frictional force between the shaft and the mass was small. To fix the spring, it is press-fitted into a bushing that is fixed to the ground. The mass and the spring were not fixed. The mass is in a free oscillation state when it is in contact with the spring. When the mass is separated from the spring, the mass is in a free fall state. Therefore, this equipment can represent the dynamics of a PWL system that moves between two states with one degree of freedom and produces the behavior of a spring-mass-hopping system.

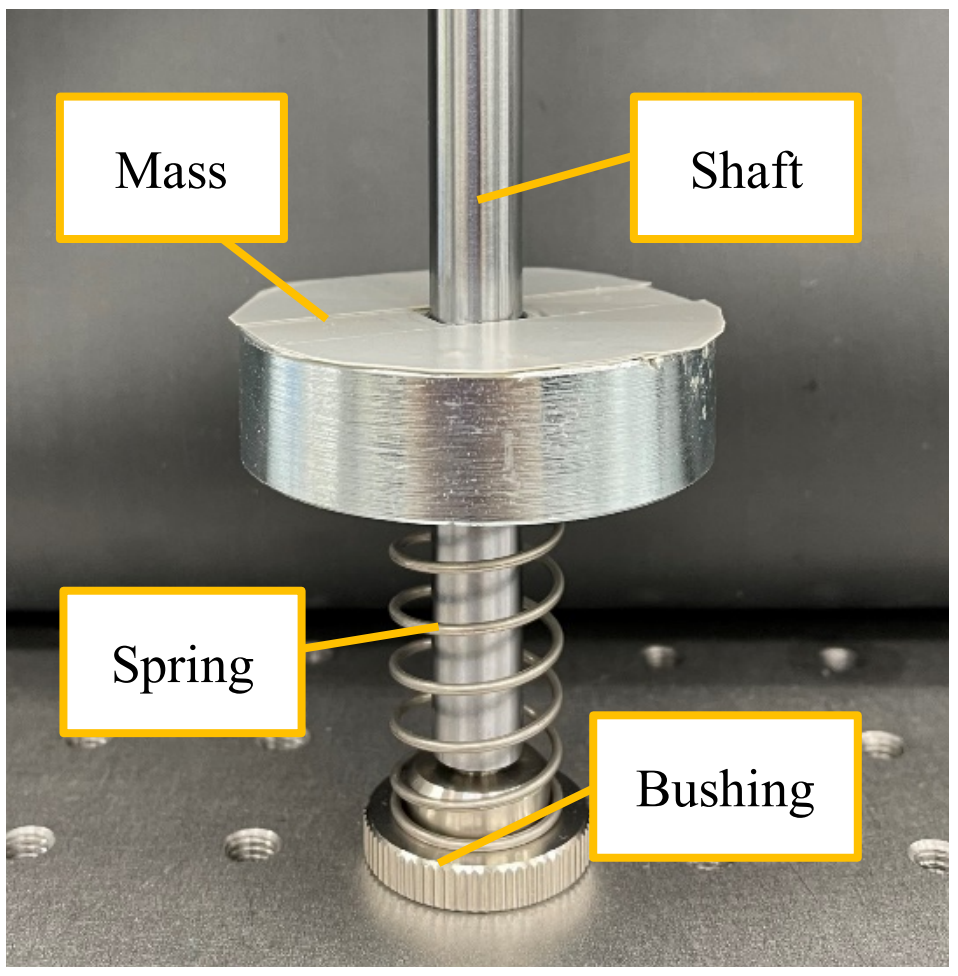

Figure 10: Experimental setup of the mass-spring-hopping system.

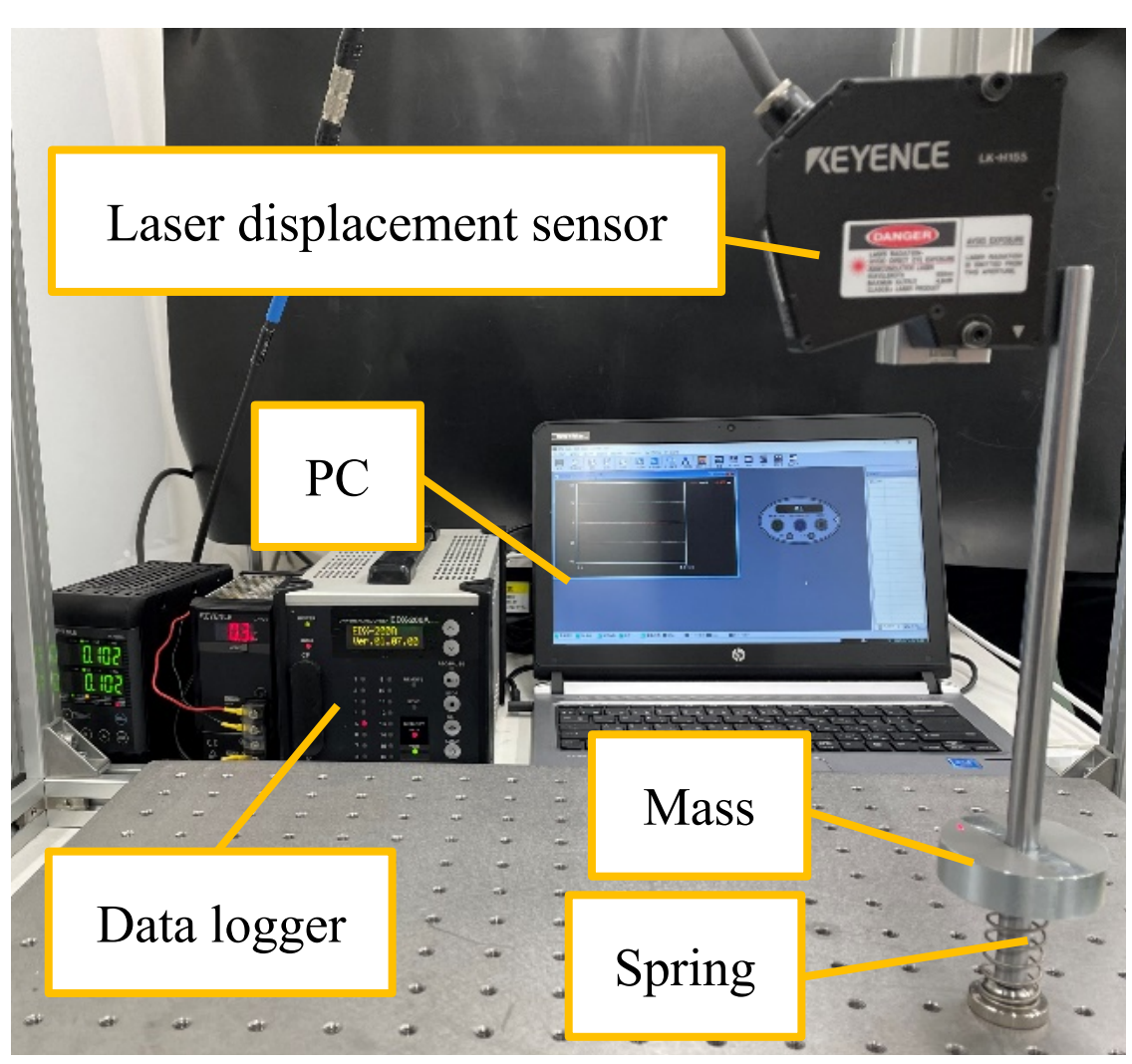

Figure 11: Experimental equipment and measuring devices.

## 4.2 Parameter identification

To identify the various parameters of the fabricated experimental equipment, the gap and the spring constant were identified from data measurements at the equilibrium position. The gap $L$ of this equipment is the distance from the static equilibrium position to the natural length of the spring. The position of the mass was measured with a laser displacement sensor to identify the gap. The spring constant was calculated from Eq. (21) based on the obtained gap. The expected governing equation for the mass shown in Fig. 10 are assumed to be as follows:

$$m\ddot{x} + c_1\dot{x} + c_2\dot{x}_{\mathrm{pwl}} + k\min\{0, x - L\} = -mg \tag{22}$$

where $m, c_1, c_2,$ and $k$ are mass, damping coefficients, and spring constant. $c_2\dot{x}_{\mathrm{pwl}}$ in Eq. (22) represents the damping force generated when the mass contacts the spring, and defined as follows:

$$\dot{x}_{\mathrm{pwl}} = -\dot{x}\min\{0, \mathrm{sgn}(x - L)\}, \tag{23}$$

where $\mathrm{sgn}(x)$=1 for $x > 0$, 0 for $x = 0$, or $-1$ for $x < 0$. The mass, spring constant, and gravitational acceleration in Eq. (22) were fixed, and a least squares method was used on the measured data to the find the damping coefficients. The obtained parameters of Eq. (22) are shown in Table 5.

Table 5: Coefficients of spring-mass-damper system.

| $m$[kg] | $c_2$[Ns/m] | $c_2$[Ns/m] | $k$[N/m] | $g$[m/s$^2$] | $L$[m] |
|---|---|---|---|---|---|
| 0.2088 | 0.333 | 1.404 | 494.526 | 9.810 | 4.142$\times 10^{-3}$ |

Table 6: Measuring instruments.

| Equipment | Manufacturer | Model number |
|---|---|---|
| Laser displacement sensor | KEYENCE Corp. | LK-H155 |
| Display panel | KEYENCE Corp. | LK-HD500 |
| Power supply | KEYENCE Corp. | CA-U4 |
| Data logger | KYOWA Co., Ltd. | EDX-200A |

## 4.3 Analysis conditions

The measurement system is shown in Fig. 11. A laser displacement sensor was used to measure the mass moving along the shaft. Velocity and acceleration were obtained by differentiating the measured displacements numerically. The values measured by the laser displacement sensor were stored in a PC through a data logger. Table 6 lists the measuring instruments used to measure the displacement of the mass. The conditions for gap identification of the experimental equipment using the proposed method are described as follows. The available data were displacement, velocity, and acceleration. The sampling frequency of the data was 10kHz, and the data were collected in the interval from $t = 0$ to $t = 3$ [s]. After measuring the displacement using the laser displacement sensor, a low-pass filter was applied to remove high-frequency noise. The `signal.sosfiltfilt` function in Python's Scipy library was used as the filter. The cutoff frequency was set to 140 Hz and the filter order was set to 2. The cutoff frequency was set to about 20 times the natural frequency of the mass when the mass is attached to the spring. The order of the filter was set to the minimum value to suppress rapid damping so as not to impair the frequency response. After applying the low-pass filter to the displacement, velocity was computed by differentiating the displacement by the central difference method. Similarly, acceleration was computed by differentiating the velocity also by the central difference method. The threshold for applying the sparse regression to the measurement data was set to $\lambda = 0.95$. To limit the generation of extra terms caused by the influence of measurement noise in the data, the threshold was set higher than in the numerical analysis. Note that the threshold should not exceed 1.0, because $\dot{x} = 1v$ when the governing equations are transformed into a state space model. Therefore, setting the threshold above 1.0 makes it impossible to construct the equation as a state space model. The library $\mathbf{\Theta}$ includes polynomial terms of up to the order of $O = 3$, and the number of candidate max functions was set to $n = 10$. The gap $L_i$ for each min function was chosen to be $L_j = (j-1)L_0,\ L_0 = 1.0\ \text{[mm]},\ j = 1, \cdots, n$. The setup is assumed to have PWL damping coming from the spring. Therefore, once the gap is derived by the regression, a PWL velocity term $\dot{x}_{\text{pwl}}$ is generated whose value becomes valid when the displacement is below the gap. The regression was then performed again with $\dot{x}_{\text{pwl}}$ included in the library to derive the equations. The PWL velocity term can be calculated as in Eq. (23).

### 4.4 Results

In this section, we obtain an expression for Eq. (22) in the form of state space model and identify the gap from the obtained expression. The state space model in Eq. (22) can be transformed as follows:

$$\begin{aligned} \dot{v} &= -g - \frac{c_1}{m}v - \frac{c_2}{m}\dot{x}_{\text{pwl}} - \frac{k}{m}\min\{0, x - L\} \\ \dot{x} &= v \end{aligned} \tag{24}$$

Table 7 shows the coefficients obtained by the proposed method and the identified gaps. Note that the column corresponding to $\dot{v}$ with Reference label shows the coefficients of the equations obtained by a standard least squares method with fixed values for the gravitational acceleration, spring constant, and measured gap. From the obtained equation coefficients, the terms that do not contribute to the dynamics does not appear in the obtained equation, which means the unnecessary terms are eliminated. The damping term corresponding to $v$ was identified to be zero. Examining the value of the Reference from Table 7, the corresponding damping coefficient is $0.333[\text{Ns/m/kg}]$, which was below the threshold. This resulted in eliminating the damping term. This was necessary because the threshold value was set to 0.95 to reduce the effects of noise associated with measurement errors. As a result, instead of the damping term being zero, the contribution of the PWL damping term has increased compared to the reference value. The coefficient of the constant term representing the value of the gravitational acceleration is $9.833[\text{m/s}^2]$. Assuming that the actual value is $9.810[\text{m/s}^2]$, the value is obtained with an error of $0.234\%$. The equivalent PWL stiffness was obtained to be about $2419[\mathrm{N/m\,/kg}]$, with an error of $2.146\%$ compared to the measured value. The equivalent gap value was obtained as $4.066\times 10^{-3}[\mathrm{m}]$, which results in the error of $1.834\%$ compared to the measured value. This indicates that the equivalent spring constant and the gap can be obtained quite accurately by the proposed method even with the real measured data. Figure 12 shows the measured data and the data obtained by fitting using the obtained model. Good agreement is obtained for both velocity and acceleration.

To evaluate the validity of the obtained model, the model accuracy and its complexity are discussed. When deriving a model from data, it is important to balance the model accuracy and its complexity [5,6,8]. Model complexity is the number of terms used to represent the original data. The more terms in the model, the better fit to the data used for training. However, fitting to noise and extrapolation can lead to completely different behavior from the one obtained with less terms. This phenomenon is called overfitting and should be avoided in machine learning. Here, the Akaike Information Criterion (AIC) [22] is considered as a criterion to examine the balance between the two indicators. The AIC can be calculated as follows:

$$\mathrm{AIC} = -2\ln H + 2K \tag{25}$$

where $H$ and $K$ are the maximum likelihood and the number of parameters, respectively. From the number of data $n$ and mean squared error $s$, $-2\ln H = ns$. $K$ is the number of parameters of the obtained model. As shown in Eq. (25), the higher the accuracy of the model and the smaller the number of parameters, the smaller the value of AIC becomes. The AIC values for the Reference and the identified results are shown in Table 7. Comparing the AIC of the reference equation with that obtained by the proposed method, there is no significant difference between them. Therefore, although the equations were not in perfect agreement, the proposed method was able to derive a model with the same level of accuracy and simplicity as the ideal equation.

Table 7: Comparison of ideal equation and the equation identified by the proposed method.

| | Reference | | Identified | |
|---|---|---|---|---|
| $\boldsymbol{\Theta}$ | $\dot{v}$ | $\dot{x}$ | $\dot{v}$ | $\dot{x}$ |
| $1$ | 9.810 | – | 9.833 | 0.000 |
| $x$ | – | – | 0.000 | 0.000 |
| $v$ | 0.333 | 1.000 | 0.000 | 1.000 |
| $x^2$ | – | – | 0.000 | 0.000 |
| $xv$ | – | – | 0.000 | 0.000 |
| $v^2$ | – | – | 0.000 | 0.000 |
| $x^3$ | – | – | 0.000 | 0.000 |
| $x^2v$ | – | – | 0.000 | 0.000 |
| $xv^2$ | – | – | 0.000 | 0.000 |
| $v^3$ | – | – | 0.000 | 0.000 |
| $\dot{x}_{\mathrm{pwl}}$ | 1.416 | – | 1.799 | 0.000 |
| $\min\{0, x - L\}$ | 2368.421 | – | – | – |
| $\min\{0, x - L_1\}$ | – | – | 543.215 | 0.000 |
| $\min\{0, x - L_2\}$ | – | – | -211.847 | 0.000 |
| $\min\{0, x - L_3\}$ | – | – | 289.329 | 0.000 |
| $\min\{0, x - L_4\}$ | – | – | 409.208 | 0.000 |
| $\min\{0, x - L_5\}$ | – | – | 431.038 | 0.000 |
| $\min\{0, x - L_6\}$ | – | – | 606.264 | 0.000 |
| $\min\{0, x - L_7\}$ | – | – | 17.963 | 0.000 |
| $\min\{0, x - L_8\}$ | – | – | -137.791 | 0.000 |
| $\min\{0, x - L_9\}$ | – | – | -88.475 | 0.000 |
| $\min\{0, x - L_{10}\}$ | – | – | 560.348 | 0.000 |
| $k_{eq}/m$[N/m/kg] | 2368.421 | | 2419.252 | |
| $L_{eq}$[m] | $4.142\times 10^{-3}$ | | $4.066\times 10^{-3}$ | |
| ln(AIC) | 11.476 | – | 11.366 | – |

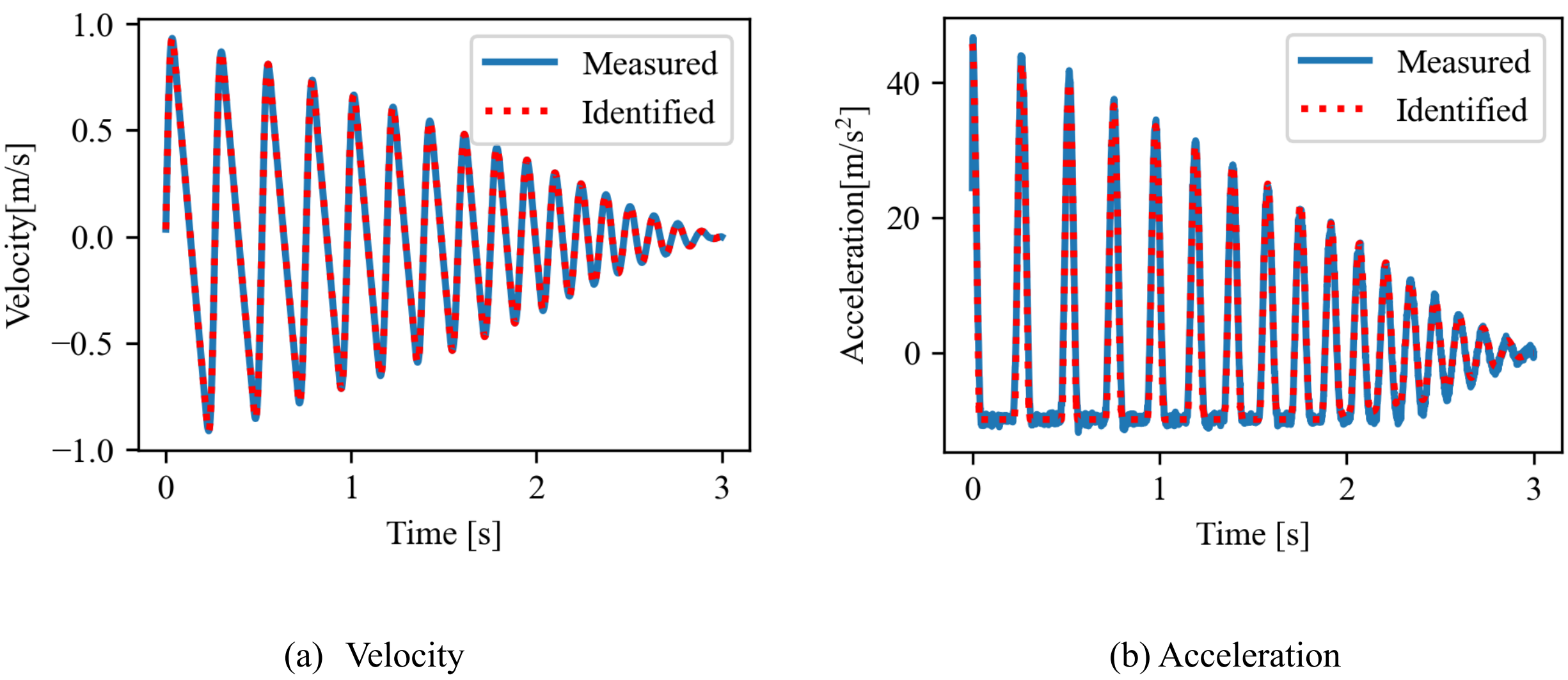

(a) Velocity (b) Acceleration

Figure 12: Validation of the proposed method on the measured data of the experimental equipment, as indicated by the velocity and acceleration.

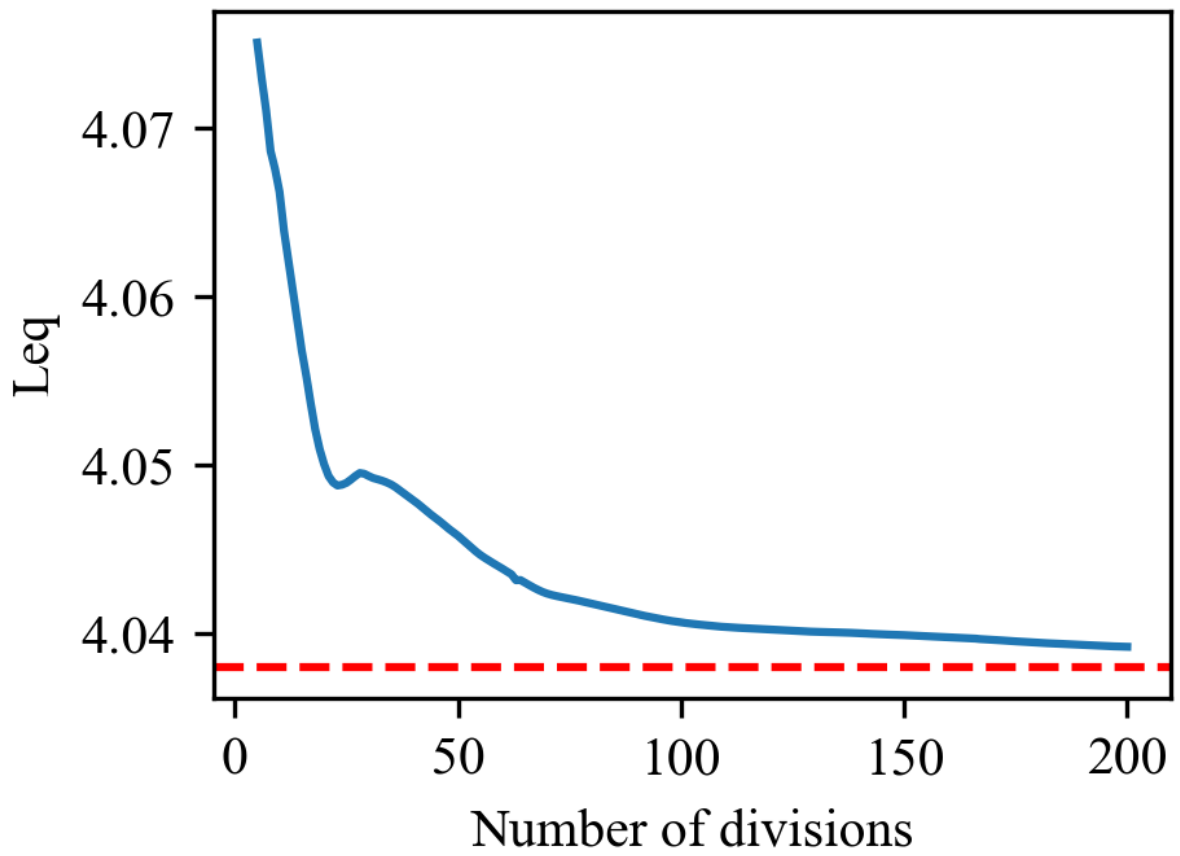

Figure 13: The convergence of the equivalent gap.

## 4.4 Convergence study

In this section, as in Section 3.3, we examine the change in the equivalent gap as the number of gap divisions is increased in the experimental analysis. The gaps of the min functions in the library are set from 0 [mm] to 9.0 [mm]. Figure 13 shows the identified equivalent gaps when the number of divisions is varied from 5 to 200. The solid line is the equivalent gap and the dashed line indicates 4.035 [mm]. As seen, the equivalent gap converges to around 4.035 [mm] as the number of divisions increases. Even though the measured gap was approximately 4.142 [mm], based on the results obtained in this section, the converged gap value found by the proposed approach is in good agreement with the real gap value with only 2.58% error. Figure 14 shows the percentage contributions of max functions used in the approximation for each number of divisions. Here, the contribution is the absolute value of the obtained coefficient, and the percentage contribution is defined to be its fraction to the sum of total contributions. Unlike the contributions in the numerical analysis in Section 3.3, almost all terms were given contributions. This means that no sparsification was performed and gaps were identified using all terms. In numerical analysis, the largest contribution was given to the PWL function close to the correct gap, and the further away from the gap, the smaller the contribution became. In the experimental analysis, however, several large peaks are observed mostly in the vicinity of the correct gap value. The contributions of those dominant peaks decrease almost monotonically with respect to the distance from the gap of the PWL function to the true gap value. Therefore, even though the convergence is not fast with respect to the number of functions in the approximations in comparison with that observed for the numerical experiments, since the contributions of the dominant min functions decrease as the distance of the corresponding $L_i$ from $L$ decreases, it shows that this approximation is convergent.

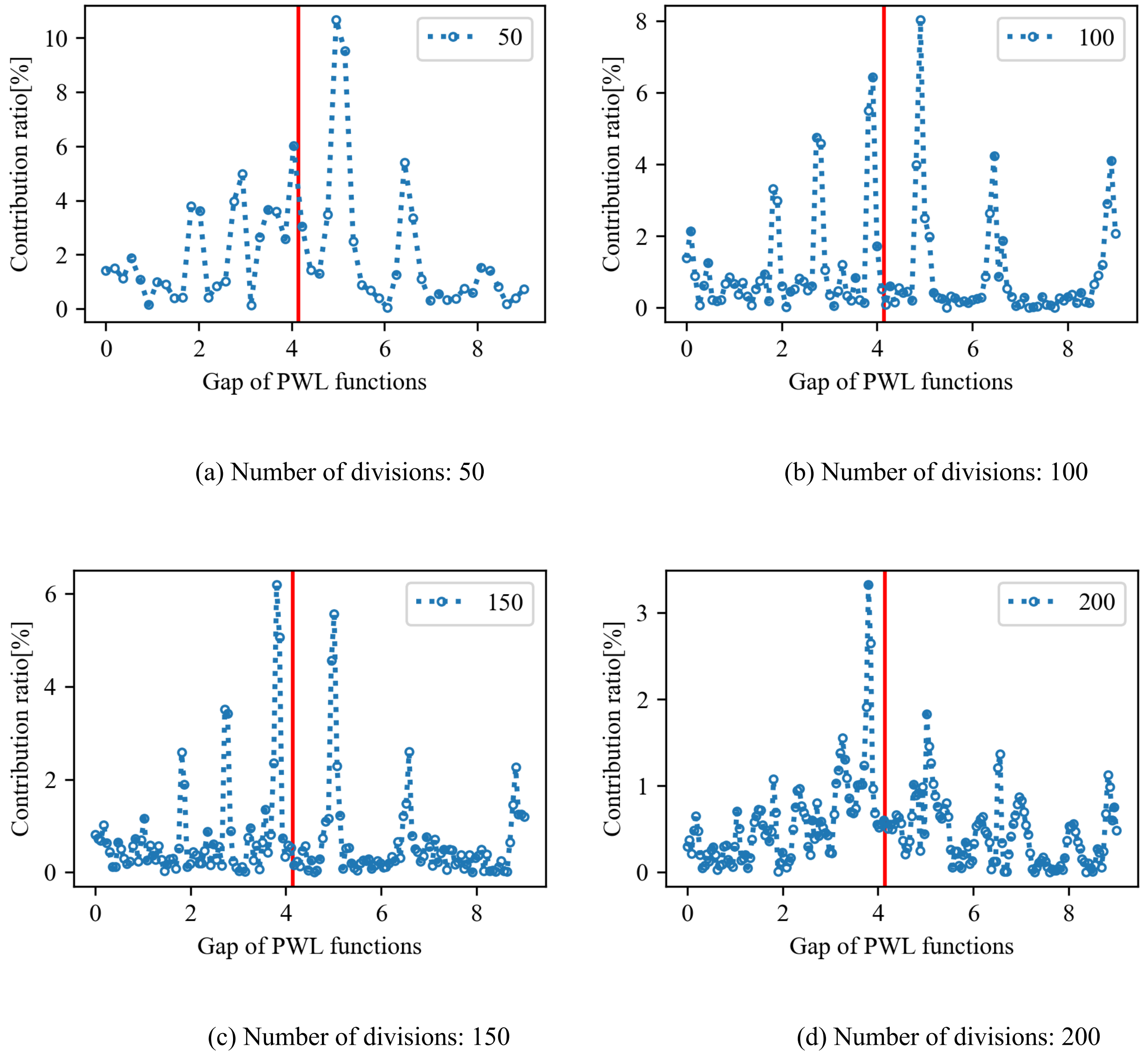

(a) Number of divisions: 50

(b) Number of divisions: 100

(c) Number of divisions: 150

(d) Number of divisions: 200

Figure 14: Contribution ratio for each number of divisions

As can be seen in the results shown in Fig. 14, all max functions considered appear in the converged results because they all had contributions larger than the set threshold value. In order to further improve the approximation result, a set of analyses have been conducted by using less number of max functions. Namely, a relative threshold that is applied only for the reduction of max functions alone is introduced. The threshold value changes at every regression, i.e.,

$$\lambda_{\mathrm{rel}} = \alpha \sum_{\mathrm{j}=1}^{n} \left|k_{\mathrm{j}}\right| \tag{26}$$

where $\alpha$ is the pre-set threshold. Note that the reason why another threshold value is introduced is because if the contributions of the max functions and other functions such as polynomials in the library are evaluated with the same threhold, functions other than the max functions may not survive during the successive application of the regression, which may result in the derivation of the governing equation that is quite different from the expected equation. The gap identification has been applied with the relative threshold Eq.(26) for α=0, 0.5, 1.0, and 2.0 with $n = 200$, and the obtained the contribution ratios are shown in Fig. 15. As expected, as the value of the relative threshold increases, the number of contributing max functions decreases. Also, the values of the identified equivalent gap $L_{eq}$ for α=0, 0.5, 1.0, and 2.0 are shown in Table 8. As can be seen in the table, since the measured gap value obtained by the laser displacement sensor was 4.142 mm, reduction in the number of max functions does not necessarily improve the identification result. This implies that the data obtained from the experimental setup requires that there be many max functions if it is fit with the model. Namely, from microscopic viewpoint, the contact between the spring and the mass may occur multiple times at different points, which also implies that the assumption that this system is a single DOF only for the identification of a single gap is not appropriate because there are indeed multiple gaps, or the gap varies on the plane perpendicular to the movement of the mass. Further investigation of this result will require thorough measurement of the distribution of the gap in the plane perpendicular to the direction of contact. However, it is beyond the scope of this paper. One thing to note is that the reduction of the number of max functions in the regression significantly reduces the computational cost. Thus, it is recommended to reduce the number of max functions if the accuracy is not greatly decreased.

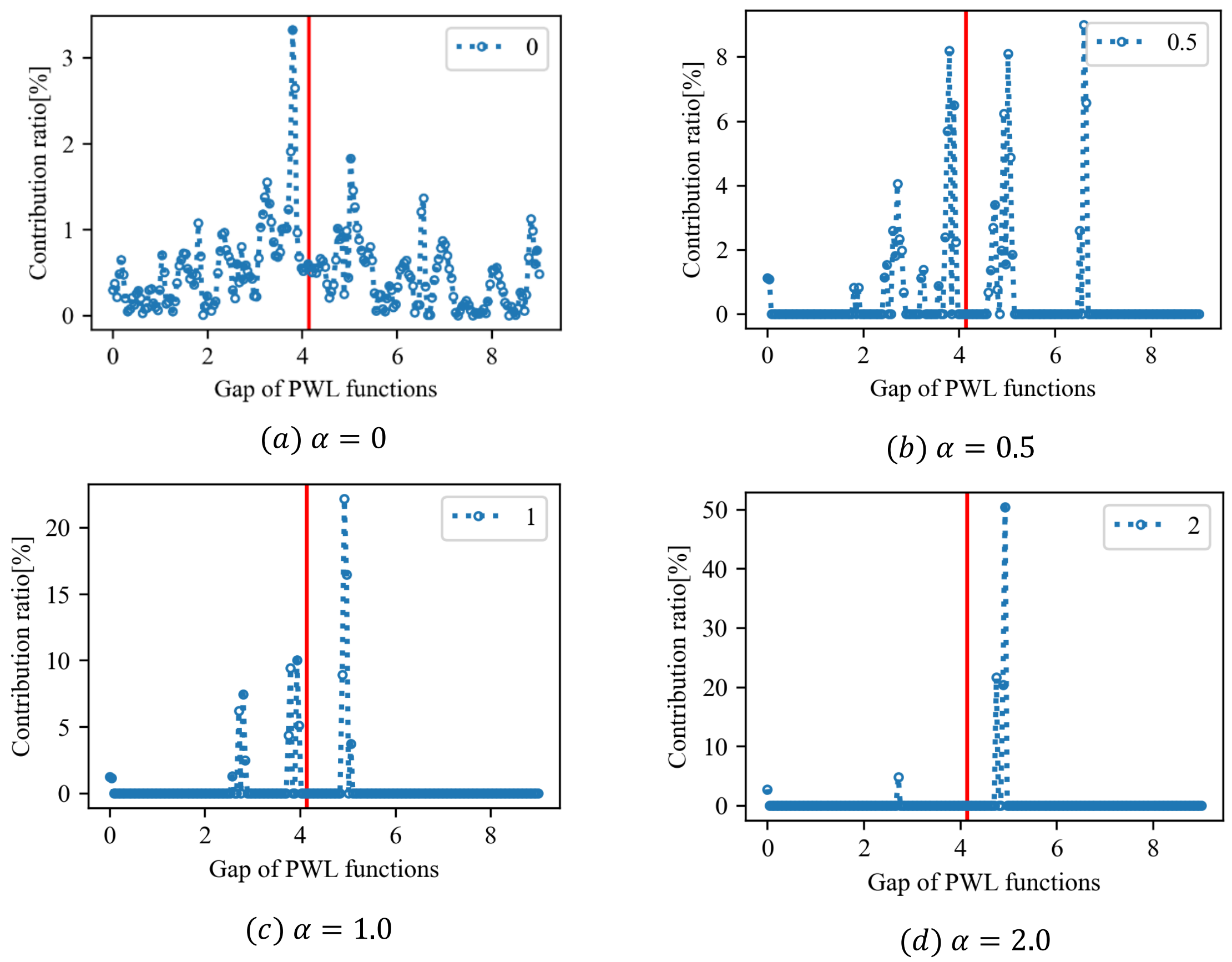

Figure 15 Contribution ratio for different relative threshold values

Table 8 Equivalent gap obtained with the relative threshold

| α [%] | $L_{eq}$[mm] | Error [%] |
|---|---|---|
| 0 | 4.039 | 2.487 |
| 0.5 | 3.951 | 4.611 |
| 1.0 | 3.684 | 11.057 |
| 2.0 | 3.708 | 10.478 |

Table 9 Equivalent gap obtained with different lower and upper bounds of $L_j$

| Lower bound [mm] | Upper bound [mm] | $L_{eq}$[mm] | Error [%] |
|---|---|---|---|
| 0.0 | 9.0 | 4.039 | 2.487 |
| 1.0 | 8.0 | 4.091 | 1.231 |
| 2.0 | 7.0 | 3.988 | 3.718 |
| 3.0 | 6.0 | 3.957 | 4.466 |
| 4.0 | 5.0 | 3.912 | 5.553 |

To see the effects of the range of values $L_j$ in the library on the identification results, another set of identifications have been conducted for different values of the lower and upper bounds of $L_j$. The number of divisions was fixed to $n = 200$. The results are shown in Table 9. As can be seen, narrowing the range between the upper and the lower bounds resulted in increasing the error. This is because by narrowing the range, the accuracy of the approximation itself decreased, which resulted in the increase in the error of the gap. This implies that we need to add max functions with gaps whose values are reasonably far from the true or expected gap value. In other words, the inclusion of max functions with gaps $L_j$ that are close to each other alone may result in low accuracy in the identified gap.

## 5 Conclusion

In this paper, a novel data-driven method for identifying a gap in PWL systems was proposed. In Section 2.4, through numerical examination, it was shown that a PWL function can be expressed using multiple PWL functions. The trajectories were in good agreement with each other with no significant errors. The obtained equations show that the sum of the coefficients is equal to the coefficient of the original PWL function. The gap can also be identified from the obtained equations. In Section 3.1 and 3.2, it was shown that the proposed method can approximate the PWL element from the measurements of the behavior of the model. Gaps were successfully identified with small errors by the proposed approach. Even when artificial measurement noise was added to data, the gap was identified accurately. Finally, in Section 4, the gap identification was conducted on the real measurements of displacement data of a mass-spring-hopping system. The proposed method was used to derive the equations and identify the gap existing in the experimental setup. The equivalent gap of the equipment was identified with a small error. The equations obtained were close to the ideal equations with slight difference in the damping term. This is due to the sparsification caused by the high threshold considering noise, which is balanced by the increased contribution of the PWL damping term. To account for the accuracy and complexity of the derived model, AIC was introduced and compared with the ideal equation. As a result, the AIC values were not significantly different between the two models. It indicates that the equations derived by the proposed method are almost equally balanced in terms of accuracy and complexity compared to the ideal equation.